\documentclass[aps,prl,amsfonts,amssymb,twocolumn,amsmath,preprintnumbers,nofootinbib,floatfix,
showpacs]{revtex4-1}
\usepackage[dvips]{graphics}
\usepackage{graphicx}
\usepackage{bm}

\begin{document}

\title{Long-range proximity effect for opposite-spin pairs in S/F heterostructures under non-equilibrium
quasiparticle distribution}
\author{I. V. Bobkova}
\affiliation{Institute of Solid State Physics, Chernogolovka,
Moscow reg., 142432 Russia}
\author{A. M. Bobkov}
\affiliation{Institute of Solid State Physics, Chernogolovka,
Moscow reg., 142432 Russia}

\date{\today}

\begin{abstract}
By now it is known that in a singlet superconductor/ferromagnet (S/F) structure
the superconducting correlations carried by opposite-spin pairs penetrate into the ferromagnet over a short distance of the order of magnetic coherence length. The long-range proximity effect (LRPE), taking place on the length scale of  the normal metal coherence length, can only be maintained by equal-spin pairs, which can be generated by magnetic inhomogeneities in the system.   
In this work we have predicted a new type of LRPE, which can take place in S/F heterostructures
under non-equilibrium conditions. The superconducting correlations in the F region are generated by
opposite-spin Cooper pairs and equal-spin pairs are not involved. The possibility for an opposite-spin
pair to penetrate into the ferromagnet over a large distance is provided by creation
of the proper non-equilibrium quasiparticle distribution there. This leads to a sharp increase 
(up to a few orders of magnitude) of the critical Josephson current
through a S/F/S junction at some values of the voltage controlling the nonequilibrium
distribution in the F interlayer. 
\end{abstract}

\maketitle

It is well known that in a singlet superconductor/ferromagnet (S/F) structure, 
the proximity effect is negligible at distances exceeding the magnetic coherence length $\xi_F=\sqrt{D/h}$ (See, for
example, \cite{buzdin05} and references therein). Here $D$ is the diffusion constant and $h$ is the exchange
energy of the ferromagnet. For the most part of the ferromagnets, which
are used for S/F heterostructures (including weak ferromagnetic alloys like CuNi \cite{ryazanov01} 
and PdNi \cite{kontos02}) this length is much
shorter than the normal metal coherence length $\xi_N=\sqrt{D/2 \pi T}$. This suppression of the proximity
effect can be understood as follows. If the magnetization direction is homogeneous in the considered system,
then the Cooper pairs, penetrating
into the nonsuperconducting part of the structure, consist of electrons with opposite spins.
Their wave function is the sum of a singlet component $| \uparrow \downarrow  \rangle -
| \downarrow \uparrow  \rangle $ and a
triplet component $| \uparrow \downarrow  \rangle +
| \downarrow \uparrow  \rangle $ with zero total spin projection $S_z=0$ on the quantization
axis. An opposite-spin Cooper pair $| \uparrow \downarrow  \rangle$ acquires the total momentum $2Q$ or $-2Q$ inside the ferromagnet as a response
to the energy difference between the two spin directions. Here $Q \propto h/v_F $,
where $v_F$ is the Fermi velocity. Combination of the two possibilities
results in the spatial oscillations of the condensate wave function $\Psi (x)$ in the ferromagnet along the direction
normal to the SF interface \cite{demler97}. $\Psi_{s} (x) \propto \cos(2Qx)$
for the singlet Cooper pair and $\Psi_{t}(x) \propto \sin(2Qx)$ for the triplet Cooper pair. 
Upon entering the nonsuperconducting region, where the pair
is not an eigenstate, it decays exponentially on the length scale $\xi_N$. 
However, due to the nonzero total momentum, acquired by the pair in the ferromagnet, there is 
an extra decay of the condensate wave function in this case, which results from the impurity averaging 
of the oscillating Cooper pair wave functions \cite{demler97}.
This extra decay takes place on the length scale $\xi_F$. It leads, in particular, 
to the significant suppression of the Josephson current through a S/F/S junction by the factor $\exp[{-d_F/\xi_F}]$,
where $d_F$ is the length of the ferromagnetic region.

The situation changes if the magnetization orientation is not fixed.
The examples are domain walls, spiral ferromagnets,
spin-active interfaces, etc. In such a system not only the singlet and triplet
$S_z=0$ components exist, but also the triplet component with $S_z
=\pm 1$ arises in the nonsuperconducting region due to the spin rotation of one of the
paired electrons. The latter component penetrates into
the ferromagnet over a large distance, which can be of the order of
$\xi_N$ in some cases. The reason is that it corresponds
to the correlations of the type $| \uparrow \uparrow \rangle$ 
with parallel spins and is not as sensitive to the exchange field as the opposite-spin correlations.
Various superconducting hybrid
structures, where this type of long-range proximity effect (LRPE) can arise, were considered in the literature
(See Refs.~\cite{bergeret05},~\cite{golubov04},~\cite{buzdin05} and references therein).
In addition, the LRPE was theoretically predicted in structures
containing domain walls \cite{volkov08,fominov07}, spin-active interfaces
\cite{asano07,eschrig03}, spiral ferromagnets
\cite{volkov06,champel08,alidoust10} and multilayered SFS systems \cite{houzet07,volkov10}.
There are several experimental works, where the long-range Josephson effect
\cite{keizer06,khaire10,anwar10} and the conductance of a spiral ferromagnet attached to two
superconductors \cite{sosnin06} were measured. 

In the present paper we show that LRPE at an S/F interface can be generated not only by equal-spin pairs with $S_z
=\pm 1$. It is also created by opposite-spin pairs with $S_z=0$ under the condition that
the appropriate non-equilibrium and spin-dependent quasiparticle distribution is produced
and maintained in the ferromagnet. At first we concentrate on the physical essence of the effect
and after that turn to the exact calculation of the Josephson current through a S/F/S junction
under the corresponding conditions.

As it was discussed above, the source of the rapid decay of an opposite-spin Cooper pair in the ferromagnet
is the impurity averaging of the rapidly oscillating pair wave function. In turn, the reason of these
rapid oscillations is the non-zero pair momentum $Q$. It is inevitably acquired by the pair of electrons,
which have the same energy (in the particular case $\varepsilon=0$) and opposite spins upon entering
the F region.
Now let us assume that the spin-dependent
quasiparticle distribution $f_{\uparrow,\downarrow}(\varepsilon)=1/[1+\exp\{({\varepsilon \pm eV})/T\}]$
is created in the ferromagnet. The energy is counted from the chemical potential of the superconductor.
Then the electrons forming a pair, which is located at the Fermi level $\varepsilon=0$ in the superconductor,
can only enter the F region with different energies $\varepsilon_{\uparrow,\downarrow}= \mp eV$, thus conserving
the total energy of the pair. As a result, the difference between the spin-up and spin-down electron momenta
is modified and in this case $Q \propto (h-eV)/v_F $. 
Therefore, the creation of appropriate spin-dependent quasiparticle distribution with $eV=h$ in the ferromagnet
makes the electrons enter the F region with different energies, but with equal (in absolute value) momenta. 
Thus, the additional rapid decay of an opposite-spin Cooper pair in the ferromagnet is absent and the decay
length can be close to $\xi_N$.

It is worth noting here that this physics is similar to some extent to the effect discussed
recently in Ref.~\cite{bobkova11}, where it was found that for a thin superconducting film the destructive effect
of the exchange field can be fully compensated by the creation of spin-dependent quasiparticle 
distribution in it. The effect reported in \cite{bobkova11} and the LRPE discussed here are two
aspects of the same problem: coexistence of singlet superconductivity and ferromagnetism under
nonequilibrium spin-dependent distribution. 

The discussed LRPE has a profound impact on the Josephson current through an S/F/S junction under
the condition of the appropriate quasiparticle distribution in the F layer. Now we turn to quantitative 
analysis of this effect. We consider a plane diffusive junction of two s-wave superconductors with the
F interlayer, which is in the parameter range $|\Delta| \ll h \ll \varepsilon_F$, where $\varepsilon_F$
is the Fermi energy of the ferromagnet. 
As we consider a non-equilibrium system, we make use of Keldysh framework of the quasiclassical theory, where
the fundamental quantity is the momentum average of the quasiclassical Green's function 
$\check g(x,\varepsilon) = \langle \check g(\bm p_f, x ,\varepsilon) \rangle_{\bm p_f}$. Here $x$
is the coordinate normal to the S/F interface and $x=0$ is the middle of the F layer. In the interlayer 
$\check g(x,\varepsilon)$ obeys the Usadel equation \cite{usadel}
\begin{equation}
\frac{D}{\pi} \partial_x (\check g \partial_x \check g)+\left[ \varepsilon \tau_3 \sigma_0 \rho_0
+ \bm h \check {\bm \sigma} \rho_0, \check g \right]=0
\label{usadel}
\enspace ,
\end{equation}
where $\tau_i$, $\sigma_i$ and $\rho_i$ are Pauli matrices in particle-hole, spin and 
Keldysh spaces, respectively. $\check {\bm \sigma}=\bm \sigma (\tau_0+\tau_3)/2+{\bm \sigma}^* (\tau_0-\tau_3)/2$
is the spin operator for a quasiparticle. Eq.~(\ref{usadel}) should be supplied with the normalization
condition $\check g^2 =-\pi^2\tau_0\sigma_0\rho_0$. The Usadel equation in the interlayer should be also 
supplemented by Kupriyanov-Lukichev boundary conditions at SF interfaces \cite{kupriyanov88}:
$\check g \partial_x \check g=-\alpha (R_F/2R_b d_F)[\check g, \check g_S]$. Here $R_b$ and $R_F$ stand for the
resistances of the S/F interface and the F interlayer, $\alpha=+1(-1)$ at the left (right) interface, 
$\check g_S$ is the value
of the Green's function at the superconducting side of the corresponding boundary.
   
It is convenient to express Keldysh part of the full Green's function via 
the retarded and advanced components and the distribution function: 
$\check g^K=\check g^R \check \varphi-\check \varphi \check g^A$. The distribution function is diagonal in particle-hole
space: $\check \varphi=\hat \varphi (\tau_0+\tau_3)/2+ \sigma_2 \hat {\tilde \varphi} \sigma_2
(\tau_0-\tau_3)/2$. The hole component $\hat {\tilde \varphi}$
of the distribution function is connected to $\hat \varphi$ by general symmetry relation \cite{serene83}
$\hat {\tilde \varphi}=-\sigma_2 \hat \varphi (-\varepsilon) \sigma_2$. In the particle-hole space 
the retarded and advanced Green's functions 
take the form $\check g^{R,A}=\hat g^{R,A} (\tau_0+\tau_3)/2+\hat f^{R,A} (\tau_1+i\tau_2)/2+\hat {\tilde f}^{R,A}
(\tau_1-i\tau_2)/2+\hat {\tilde g}^{R,A}(\tau_0-\tau_3)/2$. For the junction under consideration
the electric current through it can be written as follows
\begin{eqnarray}
j=-\frac{d}{e R_F}\int \limits_{-\infty}^{+\infty} \frac{d\varepsilon}{16\pi^2}{\rm Tr}
\left \{\left[ \hat f^R \partial_x \hat {\tilde f}^R -(\partial_x \hat f^R) \hat {\tilde f}^R  \right. \right.
\nonumber \\ 
\left. \left. -\hat f^A \partial_x \hat {\tilde f}^A + (\partial_x \hat f^A) \hat {\tilde f}^A  \right]\hat \varphi \right \}
\label{current}
\enspace .
\end{eqnarray}

We assume that the direction of the exchange field $\bm h$ is spatially homogeneous and choose the quantization axis
along the field. In this case equal-spin pairs do not occur in the interlayer. The distribution
function and the normal part $\hat g^{R,A}$ of the Green's function are diagonal matrices in spin space. 
The anomalous Green's functions only contain singlet and $S_z=0$ triplet components and can be represented 
as $\hat f^{R,A}=\hat f_d^{R,A}i \sigma_2$ and 
$\hat {\tilde f}^{R,A}=-i\sigma_2\hat {\tilde f}_d^{R,A}$, where $\hat f_d^{R,A}$ and 
$\hat {\tilde f}_d^{R,A}$ are diagonal in spin space. Further, we assume that $d_F \gg \xi_F$.
This is the most reasonable regime to demonstrate the LRPE. 

The anomalous Green's function can be easily found analytically
in the middle part of the interlayer up to the first order in the parameter $\exp[{-d_F/2\xi_F}] \ll 1$. 
It takes the form ($\sigma=\pm 1$)
\begin{equation}
(f_d^{R,A})_\sigma=\sum \limits_{\alpha = \pm 1} 4 i \pi \kappa e^{-i\alpha \chi/2} K_\sigma^{R,A} e^{-\lambda_\sigma^{R,A}
(\alpha x+d_F/2)}
\label{f_nonrez}
\end{equation}
where $\chi$ is the superconducting order parameter phase difference between the leads and 
$\lambda_\sigma^{R,A}=\sqrt{-2i \kappa (\varepsilon+\sigma h)/D}$ with $\kappa = +1 (-1)$ for
the retarded (advanced) functions. $K_\sigma^{R,A}$ should be found from the boundary conditions 
for a given S/F interface without
taking into account the influence of the other S/F interface. It is determined by the equation
\begin{eqnarray}
\lambda_\sigma^{R,A} K_\sigma^{R,A}(1-{K_\sigma^{R,A}}^2)=
\frac{1}{4\gamma_b }\left[ \sinh \Theta_S^{R,A}(1+6{K_\sigma^{R,A}}^2 \right. \nonumber \\
\left. +{K_\sigma^{R,A}}^4)-
\cosh \Theta_S^{R,A} 4 K_\sigma^{R,A}(1+{K_\sigma^{R,A}}^2)\right],~~~~~
\label{K}
\end{eqnarray}
where $\gamma_b=R_b d_F/R_F$, $\cosh \Theta_S^{R,A}$ and $\sinh
\Theta_S^{R,A}$ originate from the normal and anomalous Green's functions at the 
superconducting side of S/F interfaces. We
assume that the parameter $(R_F \xi_S/R_b d_F)(\sigma_F/\sigma_s) \ll 1$, where
$\xi_S=\sqrt{D/\Delta}$ is the superconducting coherence length in the leads, 
$\sigma_F$ and $\sigma_S$ stand for conductivities of
ferromagnetic and superconducting materials, respectively. It 
allows us to neglect the suppression of the
superconducting order parameter in the S leads near the interface
and take the Green's functions at the superconducting side of the
boundaries to be equal to their bulk values: $ \cosh \Theta_S^{R,A}=-\kappa
i\varepsilon/\sqrt{|\Delta|^2-(\varepsilon+\kappa i 0)^2}$, $\sinh \Theta_S^{R,A}=-\kappa i
|\Delta|/\sqrt{|\Delta|^2-(\varepsilon+\kappa i 0)^2}$.

However, approximation (\ref{f_nonrez})-(\ref{K}) is only valid for $|\exp{(-\lambda_\sigma^{R,A}d_F/2)}| \sim \exp[{-d_F/2\xi_F}] \ll 1 $. That is, it is not valid if $|\varepsilon +\sigma h| \lesssim \Delta$. 
If one studies equilibrium problems, this
high energy region practically does not contribute to the Josephson current and can be neglected.
At the same time, for the problem we consider it is the most important energy region for the regime
$eV \approx h$, as it is shown below. It appears that in this resonant energy region 
the solution can also be easily found analytically taking into account that for high energies 
$|\varepsilon| \sim h \gg \Delta$ the anomalous Green's function in the superconductor is small:
$\sinh \Theta_s^{R,A} \sim (\Delta/\varepsilon) \ll 1$. Therefore, the solution for $f^{R,A}$ in the interlayer region
can be found up to the first order in this parameter. It can be also expressed by Eq.~(\ref{f_nonrez}), but
$K_{\sigma}^{R,A}$ takes the form
\begin{eqnarray}
K_{\sigma,\alpha}^{R,A}=\frac{\sinh \Theta_S^{R,A}}{4\gamma_b}\times~~~~~~~~
\nonumber \\
\frac{ \lambda_\sigma^{R,A}+
\gamma_b^{-1}+e^{i \alpha \chi-\lambda_\sigma^{R,A}d_F}(\lambda_\sigma^{R,A}-
\gamma_b^{-1}) }{(\lambda_\sigma^{R,A}+
\gamma_b^{-1})^2-e^{-2 \lambda_\sigma^{R,A}d_F}(\lambda_\sigma^{R,A}-\gamma_b^{-1})^2}
\label{K_rez}
\enspace .
\end{eqnarray}

\begin{figure}[!tbh]
  \begin{minipage}[b]{0.5\linewidth}
     \centerline{\includegraphics[clip=true,width=1.5in]{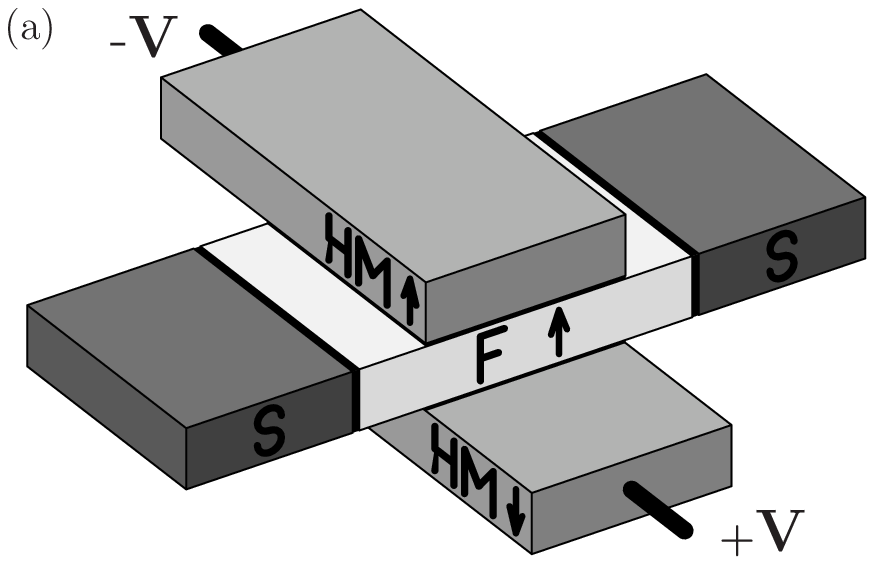}}
     \end{minipage}\hfill
   \begin{minipage}[b]{0.5\linewidth}
   \centerline{\includegraphics[clip=true,width=1.5in]{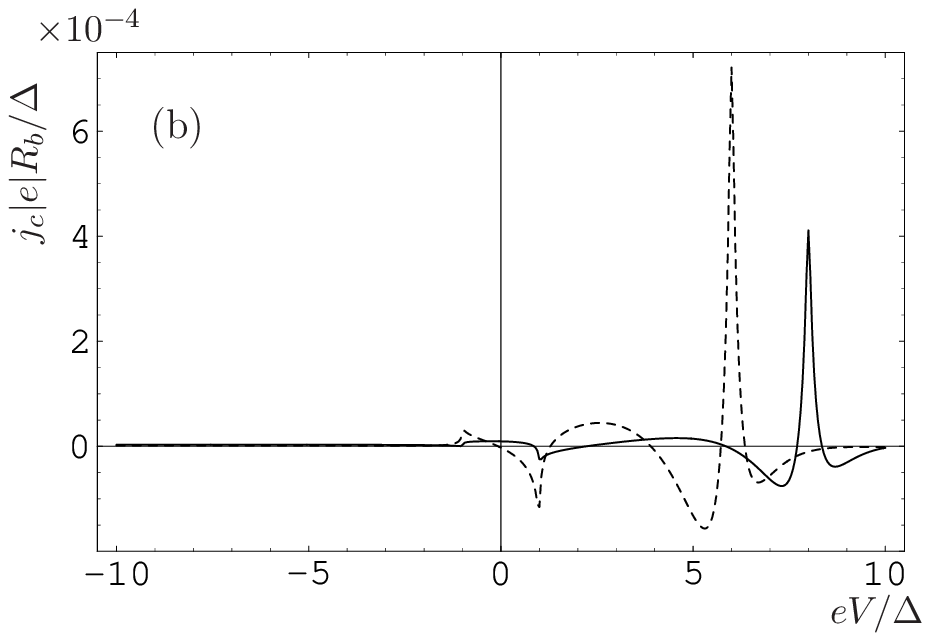}}
  \end{minipage}
\begin{minipage}[b]{0.5\linewidth}
     \centerline{\includegraphics[clip=true,width=1.5in]{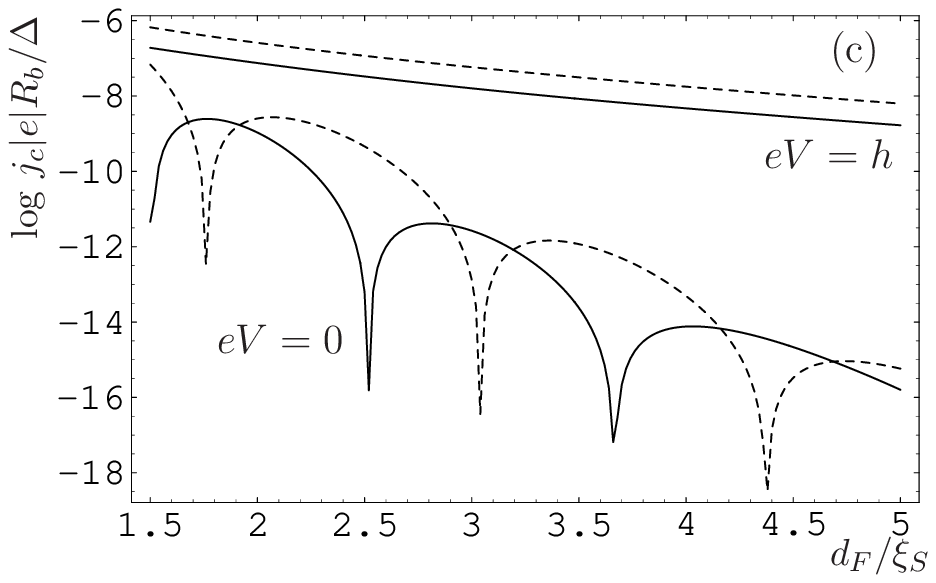}}
     \end{minipage}\hfill
    \begin{minipage}[b]{0.5\linewidth}
   \centerline{\includegraphics[clip=true,width=1.6in]{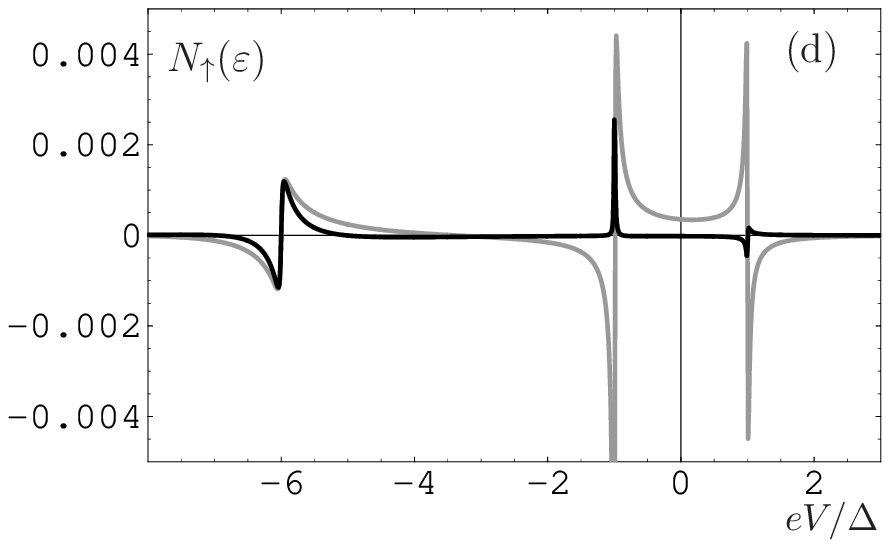}}
  \end{minipage}
   \caption{(a) Scheme of the system under consideration. (b) Critical Josephson current as a function of $eV$. $h=8\Delta$ (solid line),
$6\Delta$ (dashed line). $d_F=3 \xi_S$. (c) Critical Josephson current as a function 
of $d_F$. $h=8\Delta$ (solid line), $6\Delta$ (dashed line). (d) SCDOS $N_\uparrow$ for the spin-up subband as a function of $\varepsilon$ at $h=6\Delta$ and $d_F=2.5 \xi_S$ (black line), $1.5 \xi_S$ (gray line). 
$N_\downarrow(\varepsilon)=-N_\uparrow(-\varepsilon)$. $T=0.01\Delta$ and $\gamma_b=10$ for all the plots.}   
\label{current_V}
\end{figure}

Now let us turn to the discussion of the distribution function. In order to create the spin-dependent
quasiparticle distribution in the interlayer one can attach two additional half metal (HM) electrodes 
to the F region (see Fig.~\ref{current_V}(a)) and apply a voltage bias $2V$ between them. The magnetization 
of one of the HM's is directed along with the exchange field of the interlayer and the magnetization
of the other one is opposite. We neglect energy 
relaxation in the interlayer, that is assume that the time $\tau_{esc}$, which an electron 
spends in the F region is much less than the energy
relaxation time $\tau_\varepsilon$. Spin relaxation processes are also not taken into account. We discuss their
influence below. Then it can be calculated that
the distribution function in the film takes the form
\begin{equation}
\varphi_\sigma = \tanh \frac{\varepsilon+\sigma eV}{2T}
\label{distrib}
\enspace .
\end{equation}
In this case $\tilde \varphi_\sigma = \varphi_\sigma$.
Eq.~(\ref{distrib}) has a simple physical interpretation. For spin-up subband the main voltage 
drop occurs at one of HM, while for spin-down subband - 
at the other. As a result, the distribution functions for spin-up and spin-down electrons in the interlayer
are to be close to the equilibrium form with different electrochemical potentials.
For our special case of HM/F/HM structure the resulting chemical potential of the F region is equal 
to the chemical potential of the superconducting leads. 
Indeed, the sum of the distribution functions in the two spin subbands is symmetric 
with respect to zero energy. 
 
It is worth noting here that the distribution function has such a one-step shape
(in each of the spin subbands) due to the fact that the additional electrodes are HM: 
the electrons from spin-up (spin-down) subband can flow only to/from the top (bottom) electrode. 
In this case the LRPE effect is maximal. However, the nonequilibrium LRPE can be also observed
if one takes strong ferromagnets or even normal metals instead of HMs. We discuss these 
cases below.  

The dependence of the critical Josephson current on voltage $V$, controlling the degree of spin
imbalance in the system, is represented in Fig.~\ref{current_V}(b) for two different values of the exchange field.
It is seen that in the vicinity of $eV = h$ the critical current rises strongly. The ratio of the maximal 
current $j_{h}$ at $eV=h$ to the equilibrium critical current $j_0$ at $eV=0$ can be roughly estimated as
$j_h/j_0 \sim e^{d_F/\xi_F}(\Delta/h)^2$. It grows strongly upon increasing $d_F$ and can reach several 
orders of magnitude if $d_F$ as large as several $\xi_F$. This is the manifestation of the nonequilibrium LRPE
in the Josephson current. However, we cannot say that under the condition $eV=h$ the F layer effectively
behave as a  normal metal. The value of the exchange field practically
does not affected by the spin imbalance. From the above rough estimate 
it is seen that the the S/F/S junction under the nonequilibrium condition $eV=h$ is not equivalent 
to an equilibrium S/N/S junction of the same length because of the reducing factor $(\Delta/h)^2$.
This is originated from the fact that the superconducting correlations in the leads are suppressed by
the factor $\Delta/\varepsilon$ for large enough energies $\varepsilon \sim h$, which are important
for the LRPE.

It is worth noting here that, in addition to the sharp increase of the current at $eV=h$, the current
manifests a number of $0$-$\pi$ transitions as a function of $eV$. The region of small voltages $|eV|<\Delta$
has been studied in detail in Ref.~\cite{bobkova10}.

The dependence of the critical currents $j_h$ and $j_0$ on the junction length $d_F$ is 
plotted in Fig.~\ref{current_V}(c) in the logarithmic scale. As it is well-known \cite{buzdin05}, 
$j_0$ exhibits oscillations with a period $2\pi \xi_F$ and simultaneously decays exponentially 
on the length scale of $\xi_F$. At the same time $j_h$ does not oscillate. It decays exponentially
on the length scale of $\xi_N$. In order to study in more detail this LRPE we plot in Fig.~\ref{current_V}(d)
the supercurrent-carrying density of states $N_\sigma (\varepsilon)$ (SCDOS). This quantity represents the density of states
weighted by a factor proportional to the current that each state carries in a certain direction 
\cite{volkov95,wilheim98,yip98, heikkila02}. The full current can be represented as the integrated over energy
(and summed up over spin subbands) product of the SCDOS and the distribution function. It is seen from Fig.~\ref{current_V}(d) that the amplitude
of the SCDOS low-energy part (corresponding to $|\varepsilon| \lesssim \Delta$), which determines 
the Josephson current under equilibrium conditions, diminishes very strongly as a function of the junction 
length due to the suppression by the factor $\exp{[-d_F/\xi_F]}$. At the same time the SCDOS have sharp peaks
at energies $\varepsilon = \pm h$, which correspond to the paired states with zero total momentum 
and, therefore, are not suppressed by the factor $\exp{[-d_F/\xi_F]}$.
Under equilibrium conditions these parts of the SCDOS multiplied by the corresponding distribution function
$\tanh [\varepsilon/2T]$ give very small contribution into the current. On the contrary, shifting the argument 
$\varepsilon$ of the distribution function by $\pm eV$ for spin-up and spin-down spin-subbands one makes the peaks
to give the maximal contribution to the current.

Now we discuss briefly the influence of spin relaxation, which can take place in the interlayer,
on the Josephson current. It influences directly the distribution function: (i) reduces the height
of the main step of the distribution function $\varphi_{\uparrow,\downarrow}$ at 
$\varepsilon_{\uparrow,\downarrow}^{main}=\mp eV$ and (ii) gives rise to an additional step
of the distribution function $\varphi_{\uparrow,\downarrow}$ at 
$\varepsilon_{\uparrow,\downarrow}^{add}=\pm eV$. 
The correction to the distribution function can be roughly estimated as
$\delta \varphi_{\uparrow,\downarrow}=\mp
[(\tau_{esc}/\tau_{sf})(\varphi_\uparrow-\varphi_\downarrow)]/[(1+2\tau_{esc}/\tau_{sf})]$. 
Here $\varphi_\uparrow-\varphi_\downarrow$ is defined by Eq.~(\ref{distrib}) and 
$\tau_{sf}$ is the characteristic spin relaxation time. By looking at Fig.~\ref{current_V}(d)
it is easy to see that such modification of the distribution function does not qualitatively
modify the result for $j_h$, but only reduces its magnitude by the factor $(\tau_{sf}+\tau_{esc})/
(\tau_{sf}+2\tau_{esc})$.
The additional current peak of small height $\sim j_h\tau_{esc}/(\tau_{sf}+2\tau_{esc})$ can also
appear at $eV=-h$. 
This is the essential difference between the LRPE discussed here and the superconductivity recovered by
the nonequilibrium distribution, discussed in \cite{bobkova11}. While in the later case 
spin relaxation processes lead to the effective reduction of the coupling constant 
$\lambda \to \lambda_{eff}=\lambda(1+\tau_{esc}/\tau_{sf})^{-1}$ and, therefore, can destroy the effect
quite rapidly, the Josephson current discussed here is much more stable against their influence.  

Analogous modification of the Josephson current can be observed if one uses strong ferromagnets instead
of HM's for generation of the spin-dependent quasiparticle distribution in the interlayer. In this case the nonequilibrium distribution 
function inside the interlayer is represented by a sum of the distribution functions coming from 
the top and bottom electrodes, weighted by factors depending on the interface transparencies 
(this is a double-step structure). 
In general,
if inelastic energy relaxation can be neglected in the interlayer, the distribution function at
low temperatures manifests $n$ steps of different height at different energies $\varepsilon_n$. 
In this case instead of one
peak of maximal height at $eV=h$ the LRPE generated critical current (as a function of $V$) would
exhibit $n$ peaks of the corresponding height. Therefore,
these peaks of the critical current can provide information about the particular distribution function,
created in the interlayer. For example, if normal metals are used for additional electrodes instead of HMs,
the resulting distribution function manifests a double-step spin-independent
structure $\varphi(\varepsilon)=(1/2)\left\{\tanh[(\varepsilon-eV)/2T]+\tanh[(\varepsilon+eV)/2T]
\right\}$, measured in \cite{pothier97}. Under the nonequilibrium distribution of such type
the LRPE generated critical current would manifests two peaks of the same height $j_h/2$ at $eV=\pm h$ 
instead of one peak $j_h$ at $eV=h$, as it should be for the one-step spin-dependent distribution.
It is worth noting here that the Josephson current under the above-mentioned spin-independent
distribution has been already studied at $h \ll \Delta$ in \cite{heikkila00}. 
However, in this limit one cannot speak about the LRPE generated by
nonequilibrium distribution because there is no rapid extra decay for such extremely small exchange fields. 
 

In summary, we have predicted a new type of LRPE, which can take place in S/F heterostructures
under non-equilibrium conditions. The condensate wave function in the F region is generated by
opposite-spin Cooper pairs and equal-spin pairs are not involved. The possibility for an opposite-spin
pair to penetrate into the ferromagnet over a large distance $\sim \xi_N$ is provided by creation
of the proper non-equilibrium quasiparticle distribution there. The LRPE
can be observed as a sharp increase (up to a few orders of magnitude) of the critical Josephson current
through a S/F/S junction under the condition that the voltage controlling the nonequilibrium
distribution in the F interlayer is adjusted appropriately.

{\it Acknowledgments.} The authors are grateful to V.V. Ryazanov for useful discussions.



\end{document}